# New theoretical method for calculating the radiative association cross section of a triatomic molecule: Application to $N_2$-$H^-$

T. Stoecklin[1*], F. Lique[2] and M. Hochlaf[3]

*1 Institut des Sciences Moléculaire, UMR5255-CNRS,*
*Université de Bordeaux, 351 cours de la Libération,*
*33405 Talence Cedex, France*
*2 LOMC - UMR 6294, CNRS-Université du Havre*
*25 rue Philippe Lebon, BP 540, 76058, Le Havre, France*
*3 Université Paris-Est, Laboratoire Modélisation et Simulation Multi Echelle,*
*MSME UMR 8208 CNRS, 5 bd Descartes, 77454 Marne-la-Vallée, France.*

**Abstract**

We present a new theoretical method to treat the atom diatom radiative association within a time independent approach. This method is an adaptation of the driven equations method developed for photodissociation. The bound states energies and wave functions of the molecule are calculated exactly and used to propagate the overlap with the initial scattering wave function. In the second part of this paper, this approach is applied to the radiative association of the $N_2H^-$ anion. The main features of the radiative association cross sections are analysed and the magnitude of the calculated rate coefficient at 10 Kelvin is used to discuss the existence of the $N_2H^-$ in the interstellar medium which could be used as a tracer of both $N_2$ and $H^-$.

* Corresponding author:  t.stoecklin@ism.u-bordeaux1.fr

## 1. Introduction

Radiative association (RA) in ion-molecule collisions is considered to be an important step of the synthesis of polyatomic species in interstellar clouds [1,2]. The cosmic rays ionise the atoms and the molecules which can then recombine by radiative association to produce new molecules. As a matter of fact the very low density of atoms and molecules in even the densest clouds means that stabilization of any collision complexes must occur by the emission of a photon, rather than by collisions involving a third body. About 14 positive ions have been detected so far in the Interstellar Medium as well as several carbon chain anions. Because of its abundance, the $H_3^+$ ion is the most documented but the radiative association of the most abundant positive ions with $H_2$ are also considered in the astrochemical models. Conversely, the detection of stable negative ions in dense molecular clouds was a real and recent surprise and they are not yet taken into account in the models. For a long time, the radiative association of diatomic molecules has been the object of many theoretical [3] studies while the radiative association of triatomic molecules has up to very recently not received much interest compared to photodissociation which is the reverse process. The main reason of this lack of studies is the experimental difficulty to measure these cross sections which limited the experimental effort to the systems [4] expected to be the most important for the chemistry of interstellar clouds. On the theoretical side for many years all studies were based on statistical approximation and differed very often from the experimental estimates by several orders of magnitude. The most successful of these methods [5,6] which includes tunnelling gives rate coefficient values which are expected to be one order of magnitude accurate. The first state to state quantum calculation of the dynamics of the radiative association reaction was performed by Mrugala et al [7] in 2003 for the He-$H_2^+$ complex using the centrifugal sudden approximation. The first Close Coupling study which is even more recent was performed in 2011 by Ayouz et al [8] for the $H_3^-$ anion. This later study was motivated by the search for a probe for the presence of $H^-$ in the interstellar medium. We present here a new method to treat the atom diatom radiative association within a time independent approach. We use the formal theory of three dimensional photodissciation of a triatomic molecule which is the reverse process and was developed long ago by Band Freed and Kouri [9] and Balint Kurti and Shapiro [10]. We will more specifically present an adaptation of the driven equations method developed at the same time for photodissociation by Heather and Light [11]. In the second part of this paper, this approach is applied to the radiative association of the $N_2H^-$ anion. The bound states energies and wave functions of this anion which we calculated exactly in a recent study [12] are used to propagate the overlap with the initial scattering wave function.

The main features of the radiative association cross sections are analysed and the magnitude of the calculated rate coefficient at 10 K is used to discuss the existence of the $N_2H^-$ in the interstellar medium which could be used as a tracer of both $N_2$ and $H^-$.

## 2. Method

The method of the driven equations as initially developed for photodissociation relates in a single equation the initial and final nuclear wave functions of the colliding system. We will only present the main steps of this method which for the differences between radiative association and photodissociation are leading to new formulation and we will use greek index for the bound states and latine letters for the scattering states. A first simplification of the application of this approach to radiative association is due to the fact that very often a single electronic PES participates to the radiative association process while at least two different electronic surfaces are involved for photodissociation. We then write the total wave function of the system like a superposition of the initial and final states:

$$\Psi_{Total} = \zeta_e \Psi_i(R,r,\gamma)|0\rangle + \zeta_e \Psi_f^\alpha(R,r,\gamma)|1\rangle \qquad (1)$$

Where $\zeta_e$ and $|0\rangle, |1\rangle$ are respectively the electronic state of the triatomic system and the photon states. A second simplification appears in this first equation as we take the same space fixed Jacobi coordinates $(R,r,\gamma)$ to describe the initial $\Psi_i$ and final $\Psi_f^\alpha$ nuclear wave functions whereas Heather and Light used different coordinates for photodissociation. Here the scattering state associated with the collision of $N_2$ and $H^-$ is the initial state $\Psi_i$ of the triatomic system whereas it would be the final state in a photodissociation process. After operating on the left by $\langle 0|\zeta_e^*$ and $\langle 1|\zeta_e^*$ and integrating one obtains a system of coupled equations:

$$[\hat{H} - E]\Psi_i(R,r,\gamma) = \mu(R,r,\gamma)\Psi_f(R,r,\gamma) \qquad (2)$$

$$[\hat{H} - E]\Psi f(R,r,\gamma) = \mu(R,r,\gamma)\Psi_i(R,r,\gamma) \qquad (3)$$

Where $\mu(R,r,\gamma)$ is the dipole moment function of the lowest electronic state of the triatomic system and $\hat{H}$ is the atom-diatom inelastic scattering Hamiltonian in Jacobi coordinates:

$$\hat{H} = \left[ -\frac{\hbar^2}{2\mu_{A-BC}} \left( \frac{1}{R} \left( \frac{\partial^2}{\partial R^2} \right) R \right) + \frac{\hat{l}^2}{2\mu_{A-BC} R^2} - \frac{\hbar^2}{2\mu_{BC}} \left( \frac{1}{r} \left( \frac{\partial^2}{\partial r^2} \right) r \right) + \frac{\hat{j}^2}{2\mu_{BC} r^2} + V(R,r,\gamma) \right] \quad (4)$$

with $\hat{j}$ and $\hat{l} = \hat{J} - \hat{j}$ being respectively the operators associated with the rotational angular momentum of the diatomic molecules and with the relative angular momentum. As we know the final bound state wave functions we can ignore the second equation and look for the initial state wave function $\Psi_i(R,r,\gamma)$ solution of equation (1). This is exactly the reverse of photodissociation which for the initial state is known and the final state is looked for. The computation of the bound states energies $\varepsilon_\alpha^f$ and wave functions $\Psi_f^\alpha(R,r,\gamma)$ of the triatomic molecule in spaced fixed coordinates which is the next step of the calculation was detailed in our first paper dedicated to $N_2H^-$ [12] and will not be described again.

Both the initial and final wave functions are now expended in a basis set $\varphi_i(r)$ describing the vibration of the diatomic molecule $N_2$ and in a coupled space fixed angular basis set describing both the rotation of $N_2$ and the relative movement of $N_2$ towards $H^-$.

$$Y_{jl}^{JM}(\hat{R},\hat{r}) = \sum_{m_j,m_l} \langle jm_j l,m_l \| JM \rangle y_j^{m_j}(\hat{r}) y_l^{m_l}(\hat{R}) \quad (5)$$

$$\Psi_i(R,r,\gamma) = \frac{1}{Rr} \sum_{v,j,l} \chi_{v,j,l}^{JM}(R) \varphi_{vj}(r) Y_{jl}^{JM}(\hat{R},\hat{r}) \quad (6)$$

$$\Psi_f^\alpha(R,r,\gamma) = \frac{1}{Rr} \sum_{v,j,l} \omega_{v,j,l}^{\alpha J'M'}(R) \varphi_{vj}(r) Y_{jl}^{JM}(\hat{R},\hat{r}) \quad (7)$$

We obtain for the radial part of the scattering wave function associated with the movement of $H^-$ relative to $N_2$ the usual Close Coupled inelastic equations but with a non zero right hand side describing the dipolar coupling with the final bound states:

$$\left[ \frac{d^2}{dR^2} - \frac{l(l+1)}{R^2} + k_{vj}^2(E) - U_{vjl}^{v'j'l'}(R) \right] \chi_{vjl}^{v'j'l'}(R) = \lambda_{vjl}^\alpha(R) \quad (8)$$

Where

$$\lambda_{vjl}^\alpha(R) = -2\mu_{H-N_2} \int dr d\hat{r} d\hat{R} \varphi_v(r) Y_{jl}^{JM}(\hat{R},\hat{r}) \mu(\vec{R},\vec{r}) \Psi_f^\alpha(\hat{R},\hat{r}) \quad (9)$$

is called the driving term and is a real function while $\chi_{vjl}^{v'j'l'}(R)$ is complex:

$$\chi_{vjl}^{v'j'l'}(R) = \chi_{vjl}^{v'j'l'}(R)\Big|_R + i \chi_{vjl}^{v'j'l'}(R)\Big|_I \quad (10)$$

The imaginary part of the initial wave function appears then to be a simple inelastic scattering wave function as it follows the usual Close Coupled equations:

$$\left[\frac{d^2}{dR^2} - \frac{l(l+1)}{R^2} + k_{vj}^2(E) - U_{vjl}^{v'j'l'}(R)\right]\chi_{vjl}^{v'j'l'}(R)\bigg|_I = 0 \qquad (11)$$

While the real part is solution of a system of equations involving the driving term.

$$\left[\frac{d^2}{dR^2} - \frac{l(l+1)}{R^2} + k_{vj}^2(E) - U_{vjl}^{v'j'l'}(R)\right]\chi_{vjl}^{v'j'l'}(R)\bigg|_R = \lambda_{vjl}^{\alpha}(R) \qquad (12)$$

When compared to ours, the formalism of the Franck-Condon approach chosen by Mrugala et al [7] or Ayouz et al [8] appear to be quite different as they also use different numerical methods. The approach of Mrugala et al is not strictly equivalent to ours as they use the Centrifugal sudden approximation, while the Close Coupling approach used by Ayouz et al is expected to be equivalent to ours as demonstrated by by Takatsuka and Gordon [13]. In general the dipole moment surface has to be determined *ab initio*. In the present study dedicated to the $N_2$-$H^-$ system, we checked by comparing with *ab initio* values that a very good approximation of the dipole moment of the complex is to take it to be lying along the intermolecular coordinate:

$$\vec{\mu} \simeq \left[\frac{\mu_{H-N_2}}{m_H}\right]\vec{R} \Rightarrow \mu_m \simeq R\left[\frac{\mu_{H-N_2}}{m_H}\right]C_1^m(\hat{R}) \qquad (13)$$

where $C_1^m(\hat{R})$ is a spherical harmonics in the Racah normalisation. We obtain for the driving term the simple following expression:

$$\lambda_{vjl}^{\alpha}(R) = \frac{-2(\mu_{H-N_2})^2}{m_H}\delta_{v,v'}\sum_{v',j',l'}\Gamma_{v,j,l}^{v',j',l'}R\omega_{v',j',l'}^{\alpha,J',M'}(R) \qquad (14)$$

where $\Gamma_{v,j,l}^{v',j',l'}$ contains the angular part of the integral.

$$\Gamma_{v,j,l}^{v',j',l'} = (-1)^{[2J+l+l'+1]}\left[(2J+1)(2J'+1)(2l+1)(2l'+1)\right]^{\frac{1}{2}}$$
$$\delta_{j,j'}\sum_{v',j',l'}\begin{pmatrix} l & 1 & l' \\ 0 & 0 & 0 \end{pmatrix}\begin{Bmatrix} l & J & j \\ J' & l' & 1 \end{Bmatrix} \qquad (15)$$

Following the method proposed by Heather and Light we use the Magnus propagator for the propagation of the scattering wave function $\chi_{vjl}^{v'j'l'}(R)\big|_I$. This method takes advantage of the simple analytical expression of the radial part of the local scattering wave function $G_{v,j,l}^{J,M}(R)$ in the diagonal basis set of each interval $[R_n, R_{n+1}]$. The transformation $T_n$ between the asymptotic initial scattering basis set expressed in the space fixed frame and the local diagonal representation of the wave function is obtained by diagonalising the following matrix in the middle of each interval:

$$T_n^{\dagger}\left[\left\{k_{vj}^2(E)-\frac{l(l+1)}{R^2}\right\}\delta_{v',v}\delta_{j',j}\delta_{l,l'}-U_{vjl}^{v'j'l'}(R)\right]T_n = \xi_n^2 \quad for\ R = R_n + \frac{h_n}{2} \quad (16)$$

In this local diagonal representation an analytical expression of the following integral can be obtained from a Taylor expansion through quadratic terms of the driving term:

$$o_n\Big|_l^{\alpha} = \sum_{j,k}\int_{R_n}^{R_{n+1}}\left(G_n^{\dagger}\right)_{l,j}^{J,M}(R)\left(T_n^{\dagger}\right)_{j,k}\lambda_k^{\alpha}(R)dR \quad (17)$$

Where for the sake of simplicity, the index l replaces the three quantum numbers v,j,l. These contributions are accumulated along the propagation and retransformed in the initial asymptotic basis set as written in matrix form in the following expression:

$$\theta_n\Big|_i^{\alpha} = Q_{i-1}^{\dagger}\left[\left(G_n'(R_{n+1})\right)^{-1}\right]^{\dagger}\left\{\theta_{n-1}\Big|_i^{\alpha} + o_{n-1}\Big|_i^{\alpha}\right\} \quad (18)$$

Where $Q_n = T_{n-1}^{\dagger}T_n$

The propagation is performed up to a value $\bar{R} = R_{n+1}$ of the intermolecular coordinate in the asymptotic region where the transition amplitude between the initial and final state is obtained from:

$$M_i^{\alpha}(E) = \sum_{k,l} D_{k,l}\left(T_{n-1}\right)_{k,l}\theta_{n-1}\Big|_l^{\alpha} \quad (19)$$

In this expression D results from applying an incoming wave asymptotic boundary condition to the initial scattering wave function $\chi_{vjl}^{v'j'l'}(R)$ for our radiative association problem, while Heather and Light imposed an outgoing wave asymptotic boundary condition to the final scattering wave function to treat the case of photodissociation. We apply the energy density normalization of the scattering wave function [14] and find for D:

$$D_{v,j,l;v',j',l'}^{J,M} = \left[\frac{2\pi}{\mu_{H-N_2}}\right]^{\frac{1}{2}} e^{i\left[k_{vj}\bar{R}-\frac{l\pi}{2}\right]}\left[\left(k_{vj}\right)^{-\frac{1}{2}} + i\Re_n^{\dagger}\left(k_{vj}\right)^{\frac{1}{2}}\right]^{-1} \quad (20)$$

Where $e^{i\left[k_{vj}\bar{R}-\frac{l\pi}{2}\right]}$ and $\left(k_{vj}\right)^{-\frac{1}{2}}$ are diagonal matrices in the initial asymptotic basis set and $\Re_n$ is the propagated R matrix for R=$R_{n+1}$. The Cross section for radiative association between the initial scattering state and the final bound state α is then:

$$\sigma_{v,j,J}^{\alpha,J'}(E) = \frac{8\pi^2}{3k_{vj}^2}\frac{\omega_{\alpha}^3}{C^3}\sum_{M,m,M'}\left[(-1)^{[-M]}\begin{pmatrix}J & 1 & J' \\ -M & m & M'\end{pmatrix}M_{v,j,J}^{\alpha}(E)\right]^2 \quad (21)$$

Using the assumption (13), that $M_{v,j,J}^{\alpha}(E)$ is independent of m, we can rewrite this expression after summing over M, m and M':

$$\sigma_{v,j,J}^{\alpha,J'}(E) = \frac{8\pi^2}{3k_{vj}^2} \frac{\omega_\alpha^3}{C^3} \left[ M_{v,j,J}^\alpha(E) \right]^2 \quad (22)$$

The radiative association cross section for a given initial state of the diatomic molecule and for a given collision energy is:

$$\sigma_{v,j}(E) = \sum_{J,\alpha} \sigma_{v,j,J}^{\alpha,J'}(E) = \frac{8\pi^2}{3k_{vj}^2} \frac{\omega_\alpha^3}{C^3} \sum_{J,\alpha} \left[ M_{v,j,J}^\alpha(E) \right]^2 \quad (23)$$

The calculations are limited by the selection rules J'=J±1, ε'≠ε deduced from (13) and (15).

### 3. Results and discussion

We wrote a radiative association computer code which follows the steps of the calculation described in the previous paragraph. This program was checked by reproducing the results obtained by Ayouz et al for the $H_3^-$ system. We use the $N_2$ potential as well as the 3D $N_2$-$H^-$ potential and the corresponding bound state energies and wave functions presented in our first paper [12] dedicated to this system. As usual for complexes including homonuclear molecules, the two nuclear spin isomers of $N_2$ were treated separately and we took into account the states with positive energy that are located above the j = 0 state of $N_2$ but are however bound because they have no open channels for rotational predissociation. There are all the states with odd j which energy is below the j = 1 threshold as well as the states with even j which parity is equal to $(-1)^{(J+1)}$ and which energies are below the j = 2 threshold. This results in including 681 para and 668 ortho bound states in the calculations. We use for the scattering problem the same rovibrational basis set describing the $N_2$ molecule which we use to calculate the bound states of the $N_2$-$H^-$ complex. This number of ro-vibrational states of $N_2$ was varied and convergence was reached for a basis set including values of j ranging from 0 to 18 for para–$N_2$ and from 1 to 19 for ortho–$N_2$. The maximum value of the total angular momentum J giving bound states is equal to 19 both for para and ortho-$N_2$. This means that the maximum value of J included in the calculation of the radiative association cross section was J=20. The maximum propagation distance was 150 $a_0$ and convergence was checked as a function of the propagator step size.

In Figure 1-4 the driving terms $\lambda_{vjl}^\alpha(R)$ was drawn as a function of the intermolecular coordinate R for the value of the total angular momentum J=0 and for four different bound states respectively α=1, 2, 8 and 12 giving examples of behaviours associated with the lowest and the highest bound states. The different curves on these figures are associated with different basis functions vjl defined in (5). The radial extension of the driving terms follows closely the one of the associated bound state convoluted by the dipole moment and is then relatively small for the lowest bound state while it becomes quite large for the highest. On figure 5 the RA cross sections are represented as a function of collision energy for the para and ortho cases. Many resonances appear on these curves which are associated with the $N_2H^-$ quasi bound states as analysed before by Mrugala et al for $HeH_2^+$ and by Ayouz et al for $H_3^-$. The RA cross section is proportional to the integral (17) which is calculated for a given collision energy and is proportional to the overlap of the driving term with the scattering wave function. The collision energies giving scattering wave function describing a quasi bound state of $N_2H^-$ and or having a large overlap with a driving term associated with a given angular component of specific bound states will produce resonances. This simple pictorial analysis is quite satisfactory as it suggests that each resonance of the RA cross section is associated not only with a quasi bound state but also with one or several given bounds states and partial waves. This analysis can be performed by using the partial wave analysis shown in Figure 6 and 7. Each resonance is associated with a given value of J of the initial scattering state which is coupled by the dipole moment with $J\pm1$ bound states. The group of bound states participating to a resonance is then easily identified. On Figure 8 the rate coefficient obtained from Boltzmann averaging of the cross section is represented both for para and ortho $N_2$. It exhibits bumps which result from the resonances and reaches a maximum between 1 and 10 K before decreasing monotonously at higher temperature. As can be seen on this figure the decrease of the rate coefficient between 10 and 1000 K follows a $A.T^{-1}$ law as predicted by the modified thermal theory in the case of the radiative association of an atom with a linear molecule with no dipole moment [15,16]. The A coefficient used in the figure was chosen arbitrary in order to compare the slope of our curves with the law of variation predicted by the modified thermal theory. Our calculations give for the RA rate coefficient at 10 K a value of 5 $10^{-20}$ $cm^3$ $molecule^{-1}$ $s^{-1}$ which is about what we proposed using a crude model in our first paper dedicated to $N_2H^-$. It is approximately three orders of magnitude larger than the one obtained by Ayouz et al for $H_3^-$. However the abundance of $N_2$ in the ISM is expected to be roughly $10^{-4}$ the one of $H_2$. Making the same hypothesis than Ayouz et al for the principal mechanism of production and destruction of $H_3^-$ we then expect the abundance of $N_2H^-$ in the ISM to be equivalent to the one of $H_3^-$.

## 4. Conclusion

We presented a new theoretical method for calculating the RA cross section of a triatomic molecule. This method results from an adaption of the driven equations methods developed for photodissociation and is applicable to any triatomic system. It works in the space fixed frame and needs on input the bound states energies and wave functions of the triatomic complex. We applied it to the radiative association of $N_2H^-$ ion and found that it can be formed in the ISM. It would have however to be in large amount to be detected by absorption spectroscopy while it can be destroyed by cosmic rays or by neutralization with positive ions. We obtained for the RA rate coefficient a relatively large value in comparison for example with the one of $H_3^-$ but knowing that the abundance of $N_2$ is expected to be only $10^{-4}$ the one of $H_2$ we conclude that the abundance of $N_2H^-$ in the ISM should be similar to the one of $H_3^-$. We also found that the temperature dependence of the RA rate coefficient in the [10, 1000] Kelvin interval is well predicted for this system by a simple approach like the modified thermal theory.

**Acknowledgements:**
We thank Maurice Raoult, Viatcheslav Kokooline and Olivier Dulieu for valuable discussions about the comparison of the two methods.

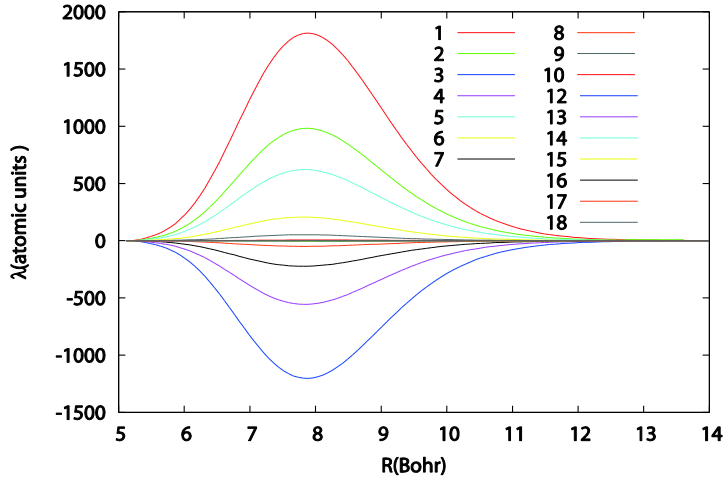

**Figure 1**: Variation as a function of the intermolecular coordinate R of the driving term associated with the lowest bound state of $N_2H^-$ ($\alpha=1$) calculated for a total angular momentum equal to 0. Each curve is associated with a given angular wave function defined in (5).

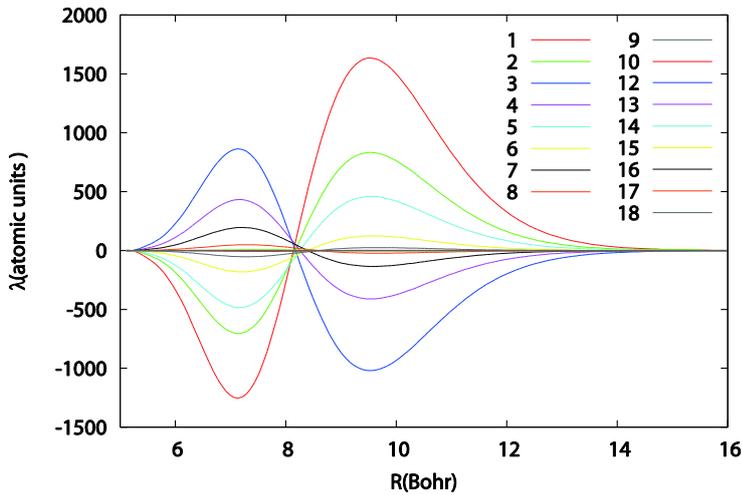

**Figure 2**: Variation as a function of the intermolecular coordinate R of the driving term associated with the bound state $\alpha=2$ of $N_2H^-$ calculated for a total angular momentum equal to 0. Each curve is associated with a given angular wave function defined in (5).

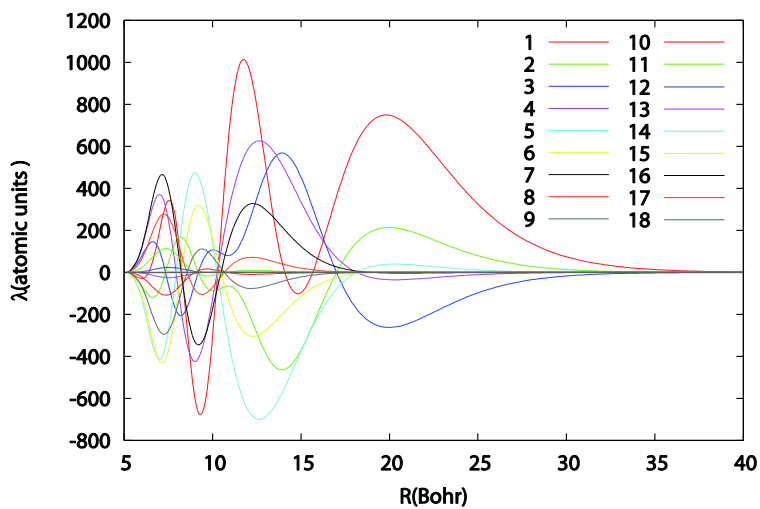

**Figure 3**: Variation as a function of the intermolecular coordinate R of the driving term associated with the bound state $\alpha=8$ of $N_2H^-$ calculated for a total angular momentum equal to 0. Each curve is associated with a given angular wave function defined in (5).

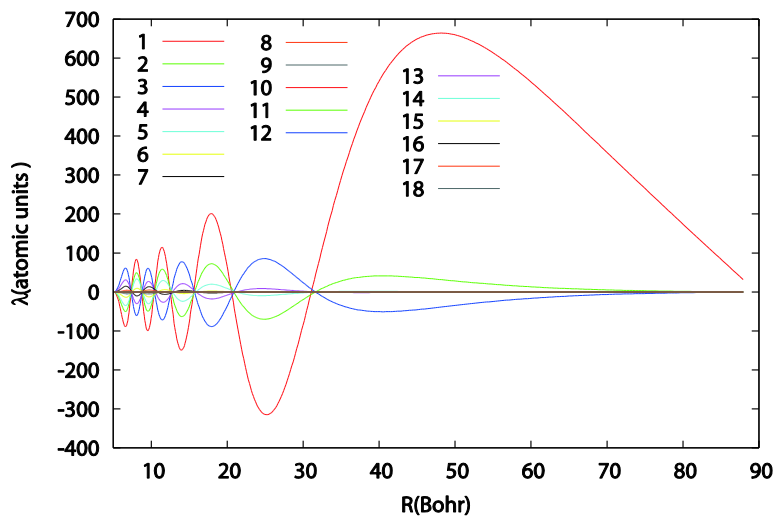

**Figure 4**: Variation as a function of the intermolecular coordinate R of the driving term associated with the bound state $\alpha=12$ of $N_2H^-$ calculated for a total angular momentum equal to 0. Each curve is associated with a given angular wave function defined in (5).

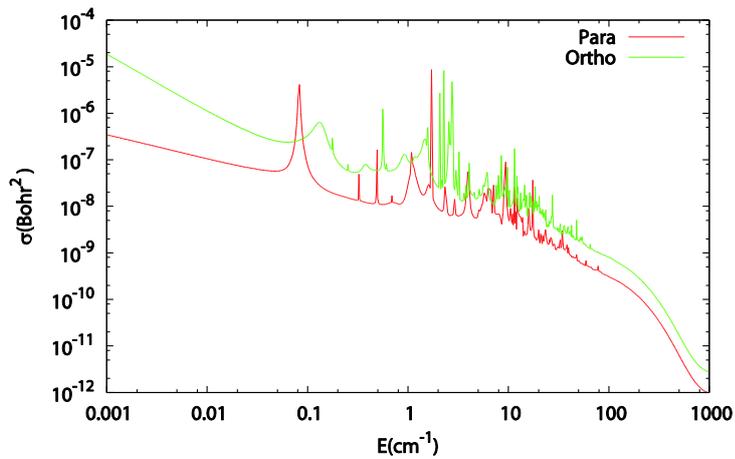

**Figure 5**: (Color online) RA cross section (in atomic units) starting either from para $N_2(v=0,j=0)$ or from ortho $N_2$ $(v=0,j=1)$ as a function of collision energy above the corresponding thresholds.

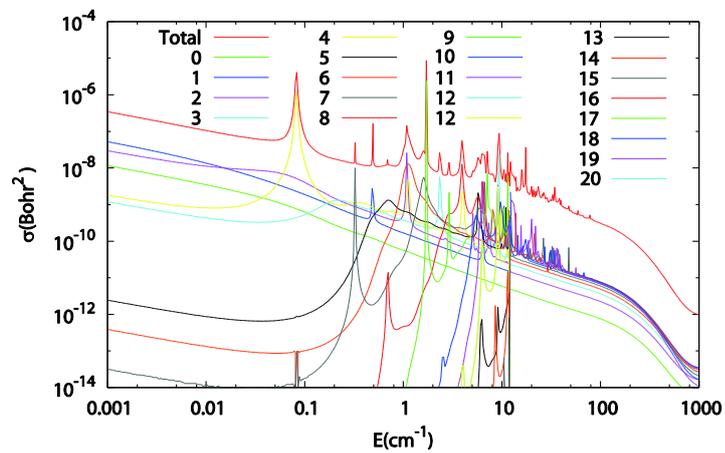

**Figure 6**: (Color online) Partial wave expansion of the RA cross section (in atomic units) starting from para $N_2(v=0,j=0)$ as a function of collision energy above threshold. The value of the total angular momentum J is indicated for each of the curves represented. The curve resulting from the summation of all the components is also reported.

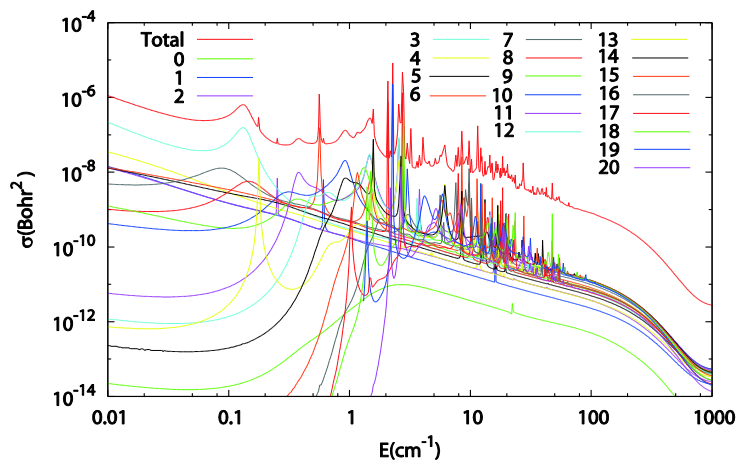

**Figure 7**: (Color online) Partial wave expansion of the RA cross section (in atomic units) starting from ortho $N_2(v=0,j=1)$ as a function of collision energy above threshold. The value of the total angular momentum J is indicated for each of the curves represented. The curve resulting from the summation of all the components is also reported.

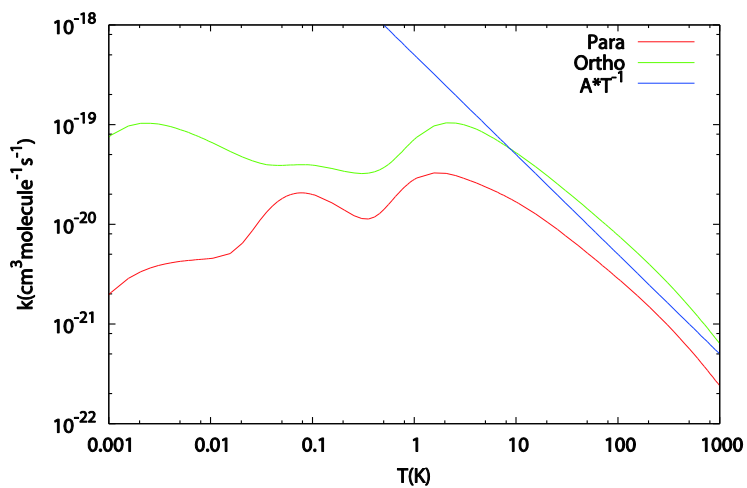

**Figure 8**: (Color online) RA rate coefficient of $N_2$-$H^-$ starting either from para $N_2(v=0,j=0)$ or from ortho $N_2$ $(v=0,j=1)$. A $T^{-1}$ law of variation is also represented.